# Incarnation of Majorana Fermions in Kitaev Quantum Spin Lattice

Seung-Hwan Do,[1,2] Sang-Youn Park,[2] Junki Yoshitake,[3] Joji Nasu,[4] Yukitoshi Motome,[3] Yong Seung Kwon,[5] D. T. Adroja,[6,7] D. J. Voneshen,[6] Kyoo Kim,[2] T.-H. Jang,[2] J.-H. Park,[2,8,9,*] Kwang-Yong Choi[1,*] and Sungdae Ji[2,8,*]

[1]*Department of Physics, Chung-Ang University, Seoul 06974, Republic of Korea.*

[2]*Max Planck POSTECH Center for Complex Phase Materials, Pohang University of Science and Technology, Pohang 37673, Republic of Korea.*

[3]*Department of Applied Physics, University of Tokyo, Bunkyo, Tokyo 113-8656, Japan.*

[4]*Department of Physics, Tokyo Institute of Technology, Meguro, Tokyo 152-8551, Japan.*

[5]*Department of Emerging Materials Science, DGIST, Daegu 42988, Republic of Korea.*

[6]*ISIS Facility, Rutherford Appleton Laboratory, Didcot OX11 0QX, United Kingdom.*

[7]*Highly Correlated Matter Research Group, Physics Department, University of Johannesburg, P.O. Box 524, Auckland Park 2006, South Africa.*

[8]*Department of Physics, Pohang University of Science and Technology, 37673, Republic of Korea.*

[9]*Division of Advanced Materials Science, Pohang University of Science and Technology, Pohang 37673, Republic of Korea.*

All the correspondence should be addressed to S.J. (e-mail: sungdae@postech.ac.kr), K.-Y.C (e-mail: kchoi@cau.ac.kr) or J.-H.P (e-mail: jhp@postech.ac.kr).

**Kitaev quantum spin liquid is a topological magnetic quantum state characterized by Majorana fermions of fractionalized spin excitations, which are identical to their own antiparticles. Here, we demonstrate emergence of Majorana fermions thermally fractionalized in the Kitaev honeycomb spin lattice α-RuCl$_3$. The specific heat data unveil the characteristic two-stage release of magnetic entropy involving localized and itinerant Majorana fermions. The inelastic neutron scattering results further corroborate these two distinct fermions by exhibiting quasielastic excitations at low energies around the Brillouin zone center and Y-shaped magnetic continuum at high energies, which are evident for the ferromagnetic Kitaev model. Our results provide an opportunity to build a unified conceptual framework of fractionalized excitations, applicable also for the quantum Hall states, superconductors, and frustrated magnets.**

Geometrical constraint often enforces the electronic states of matter to be topological quantum states such as fractional quantum Hall states, topological insulators, and Weyl semi-metals (*1-3*). In magnetism, theoretical studies predicted an entangled magnetic quantum state with topological ordering and fractionalized spin excitations, the so-called quantum spin liquid (QLS) (*4*), which holds promise for implementation in fault-tolerant quantum computers (*5*). QSL without the presence of any magnetic long-range order down to zero temperature has been predicted in geometrically frustrated magnets such as triangular, kagome, and pyrochlore lattices (*6-8*). The exchange frustration creates macroscopic degeneracy and stabilizes the topological QSL ground state. This QSL state is derived as an exact solution of the ideal two-dimensional (2D) honeycomb lattice with bond-directional Ising-type interactions ($H = J_K^\gamma S_i^\gamma S_j^\gamma; \gamma = x, y, z$) on the three distinct links (see Fig. 1A) by expressing the spin excitations in terms of non-interacting fractionalized Majorana fermions (MFs) (*9*). These fermions are their own antiparticles, distinguishable from the Dirac fermions (*10*). The elementary excitations from the Kitaev QSL are represented by localized MFs associated with static $Z_2$-fluxes and itinerant MFs (*9*) (see the cartoons in Fig. 1B). Thereby these two-types of MFs are the true physical entities that have ramifications for the observable physics (*11-15*) and potential technological applications of QSL in quantum computers (*5, 9*).

As candidates to realize a QSL, honeycomb iridates $A_2IrO_3$ ($A$ = Li, Na) with the spin-orbit coupled $J_{eff}$ = ½ $Ir^{4+}$ ($5d^5$) state (*16*) have been intensively studied. This is due to the orbital state forming three orthogonal bonds required for the bond-directional exchange interactions in the geometry (*17*). The iridates, however, cannot avoid monoclinic distortions with anisotropic Ir-Ir bonds causing a crack in the exchange frustration, and their magnetism is apparently led by antiferromagnetic (AFM) ordering (*18, 19*).

A promising candidate for the Kitaev model system could be a van der Waals ruthenate α-RuCl₃ with $J_{eff}$ = ½ $Ru^{3+}$ ($4d^5$) ions (*20, 21*). There is accumulating evidence that α-RuCl₃ host predominantly Ising-like Kitaev interactions and the ground state could be proximate to the QSL state (*22, 23*). Most crystallographic studies reported presence of the monoclinic distortions (*24, 25*), resulting in considerable contribution of the Heisenberg and asymmetric exchange interactions (*26, 27*). However, these distortions are likely due to stacking faults of the RuCl₃ layers, and even lead to multiple magnetic transitions (*25*). Recently, significant advances in the synthesis of high-quality α-RuCl₃ crystals have been achieved. These crystals are almost free from stacking faults and have a rhombohedral ($R\bar{3}$) phase, while preserving the Ising-type AFM state below 6.5 K due to non-vanishing inter-layer couplings (*28*). Importantly, this high-symmetry structure renders the isotropic Kitaev interactions ($J_K = J_K^x = J_K^y = J_K^z$) with a 94° Ru-Cl-Ru bond angle maximizing the Kitaev interaction, and the Heisenberg contribution becomes minimal (*27*). Furthermore, recent methodological progress in the quantum monte carlo (QMC) method and cluster dynamic mean-field theory (CDMFT) for thermally excited quantum states provides a route to identify the fractionalized MFs emerging from the QSL ground state (*13-15*). At very low temperature ($T < T_L$), the $Z_2$-fluxes are mostly frozen to the topologically ordered zero-temperature QSL state and the thermal energy excites only low-energy itinerant MFs (see Fig. 1B). As temperature increases across $T_L$, the fluxes fluctuate to activate the localized MFs (Kitaev paramagnet). Upon further heating, the itinerant MFs are additionally activated and the spin-spin correlation fades out across $T_H$. Finally, the system ends in the conventional paramagnetic phase well above $T_H$.

Figure 2 presents the thermodynamic signatures in the magnetic susceptibility χ(*T*) and the magnetic specific heat $C_M$ and entropy $S_M$ for fractionalized spin excitations. The static χ(*T*) of α-

RuCl$_3$ deviates from the Curie-Weiss curve below 140 K, indicating the onset of short-range spin correlations (Fig. 2A). The anomalies in χ(T) and $C_M$ at $T_N$ = 6.5 K represent occurrence of the zigzag-type AFM order (Fig. 2A,B). $C_M$ is obtained by subtracting the lattice contribution from the total specific heat ($C_P$) as described in Supplementary Materials (*29*). Besides the sharp anomaly at $T_N$, $C_M$ exhibits two broad maxima, one near $T_N$ and the other around $T_H \approx 100K$ although the low-*T* maximum feature is obscured by the AFM anomaly. As predicted in the theory (*13, 14*), the high- and low-*T* structures can be ascribed to the thermal excitations of itinerant and localized MFs, respectively. It is worth noting that $C_M$ follows a linear *T*-dependence in the intermediate range $T_N < T < T_H$, reflecting metallic-like behavior of the itinerant MFs (inset of Fig. 2B).

Rather firm evidence is provided by the two-stage release of the entropy gain $S_M(T) = \int C_M/T dT$ (Fig. 2C). The obtained $S_M$ at *T* = 200 K is 5.13 Jmol$^{-1}$K$^{-1}$, which corresponds to about 90 % of the ideal value *R*ln2 (*R*: ideal gas constant) of the spin-½ system. Upon cooling, nearly a half of the entropy is released stepwise with the plateau-like behavior at 0.46*R*ln2, signifying the two maxima of $C_M$. Indeed, $S_M(T)$ above $T_N$ well agrees with the simulated sum (red line) of two phenomenological Schotty-like functions with about an equal weight (*29*), which involve the itinerant and localized MFs in the QMC simulation. Considering the predicted temperature ratio $T_L/T_H \approx 0.03$ in the isotropic Kitaev model, the low temperature crossover $T_L$ would be somewhat lower than $T_N$ if the AFM order were absent. $S_M$ involving AFM order below $T_N$ was estimated to be 1.09 Jmol$^{-1}$K$^{-1}$, about 20% of the total entropy *R*ln2 (40% of 1/2*R*ln2) (*28*), indicating that the entropy held by the AFM order is partially released and roughly 3/5 of the frozen $Z_2$-flux is still maintained just above $T_N$.

The thermally fractionalized MFs become more visualized in microscopic and dynamic properties of the spin excitations obtained from the inelastic neutron scattering (INS) measurements. Figure 3A shows the neutron scattering function $S_{tot}(\mathbf{Q}, \omega)$ as a function of momentum transfer **Q** and energy transfer ω measured at *T* = 10 K above $T_N$ along the X-K-Γ-M-Y direction. $S_{tot}(\mathbf{Q}, \omega)$ at sufficiently low *T* can be approximated as the magnetic scattering function $S_{mag}(\mathbf{Q}, \omega)$ although weak phonon features are still observable as marked with black stars in the figure (*29*). $S_{tot}(\mathbf{Q}, \omega)$ displays an hour-glass shape spectrum centered at the Γ-point extended to about 20 meV resulting from strong low-energy (ω ≤ 6 meV) excitations around the Γ-point and high-energy (ω ≥ 10 meV) Y-shaped excitations. The similar features are reproduced in the simulated spectra in the isotropic Kitaev model with the ferromagnetic (FM) Kitaev interaction $J_K$ (< 0) by using the CDMFT + continuous-time QMC method (*15*) (see Fig. 3B). It is worth noting that the spectral center would move to the M-point for the AFM $J_K$ (> 0) (*15*). The low-energy feature represents the quasielastic responses associated with the flux excitations, and the Y-shaped **Q**-ω dependence in the high-energy region reflects the dispersive itinerant MFs extending to ω ~ |$J_K$| (*12, 15*). Both features are also clearly observable in the constant-energy cuts $S_{tot}(\mathbf{Q})$s, which also agree well with the theoretical calculations (Fig. 3C). According to the simulation, the excitation energy of the itinerant MF at the K- and M-points corresponds to Kitaev $J_K$. $S_{tot}(\mathbf{Q})$ data (Fig. 3D) are again compared with the simulated ones (Fig. 3E) in the 2D reciprocal space (Fig. 3F). The overall features are well reproduced in the simulations except the hexagram-type anisotropy of the low-energy $S_{tot}(\mathbf{Q})$, indicating that the key character of the MFs is rather robust. The hexagram-type anisotropy is considered to be induced by long-range Kitaev interactions (*30*) and/or symmetric anisotropy exchange interactions (*31*) involving the direct Ru-Ru electron hopping, both of which are not considered in the pure Kitaev model.

Figure 4A and 4B, for comparison, present the thermal evolution of the experimental and simulated $S_{mag}(\mathbf{Q}, \omega)$, respectively (*29*). At $T = 16$ K, the hour-glass shape spectrum is still maintained with minor reduction in the overall intensity. Upon heating up to $T_H \sim 100$ K (Kitaev paramagnetic phase), the low energy intensity involving the localized MFs is significantly reduced with increasing temperature while the high energy intensity from the itinerant MFs is nearly maintained although the dichotomic feature becomes smeared with the increasing thermal fluctuation. For further heating across $T_H$, the high energy intensity starts to decrease considerably. Well above $T_H$ ($T = 240$ K), $S_{mag}(\mathbf{Q}, \omega)$ only exhibits featureless low background as in conventional paramagnets. The evolution of the localized and itinerant MFs with temperature are visualized in the temperature-energy contour plots of $S_{mag}$ around Γ-point as presented in Fig. 4C (experiment) and 4D (simulation). The low-energy excitations below $\omega \approx 4$ meV appear at $T \lesssim T_H$ while the high-energy excitations extend out to $\omega \sim |J_K|$. This is also evident from the $S_{mag}(\Gamma, \omega)$ plots in Fig. 4E, which are consistent with the simulations.

The quantitative agreement between the experiment and simulation is also excellent in the INS intensities for the low and high energy excitations in an overall temperature range as shown in Fig. 4F and 4G, presenting the temperature dependences of the corresponding integrations $\int S_{mag}(\Gamma, \omega)d\omega$. Meanwhile, one also notices that the experiment somewhat deviates from the simulation below ~ 50 K only in the integration involving the low energy excitations (Fig. 4F). This is likely due to presence of the additional perturbing magnetic interactions in the real system, whose influence might be apparent in the low energy scale to be detrimental to the low-energy flux excitations at low temperature. Those perturbing interactions contribute the hexagram-type anisotropy in the low energy $S_{mag}(\mathbf{Q})$ (see Fig. 3D), which becomes isotropic above ~ 50 K as expected in the Kitaev model (*29*).

Tracing the magnetic entropy and evolution of the spin excitations as a function of temperature, energy, and momentum, we provide unambiguous evidences for thermal fractionalization to MFs of spin excitations. α-RuCl$_3$ is well described in the FM Kitaev model and is on the verge of the Kitaev QSL. The key features of the thermally fractionalized MFs predicted in the pure Kitaev model are surprisingly well reproduced in the thermodynamic and spectroscopic results, although the AFM order is developed below $T_N = 6.5$ K and additional perturbing magnetic interactions deteriorate the QSL behaviors, especially in the low energy scale. When temperature is higher than the energy scale related to the perturbing magnetic interactions, two distinct MFs predicted in the Kitaev honeycomb model are unveiled. This finding will lay a cornerstone for in-depth understanding of emergent Majorana quasiparticles in condensed matter and also possibly for future implementation in quantum computations.

**References and Notes:**

1. H. L. Stormer, D. C. Tsui, A. C. Gossard, The fractional quantum Hall effect. *Rev. Mod. Phys.* **71**, S298–S305 (1999).

2. M. Z. Hasan, C. L. Kane, Colloquium: Topological insulators. *Rev. Mod. Phys.* **82**, 3045–3067 (2010).

3. S. Y. Xu *et al.*, Discovery of a Weyl fermion semimetal and topological Fermi arcs. *Science.* **349**, 613–617 (2015).

4. L. Balents, Spin liquids in frustrated magnets. *Nature*. **464**, 199–208 (2010).

5. C. Nayak, S. H. Simon, A. Stern, M. Freedman, S. Das Sarma, Non-Abelian anyons and topological quantum computation. *Rev. Mod. Phys.* **80**, 1083–1159 (2008).

6. Y. Shen *et al.*, Evidence for a spinon Fermi surface in a triangular-lattice quantum-spin-liquid candidate. *Nature* (2016), doi:10.1038/nature20614.

7. T.-H. Han *et al.*, Fractionalized excitations in the spin-liquid state of a kagome-lattice antiferromagnet. *Nature*. **492**, 406–410 (2012).

8. S. T. Bramwell, M. J. Gingras, Spin ice state in frustrated magnetic pyrochlore materials. *Science*. **294**, 1495–1501 (2001).

9. A. Kitaev, Anyons in an exactly solved model and beyond. *Ann. Phys.* **321**, 2–111 (2006).

10. S. R. Elliott, M. Franz, Colloquium: Majorana fermions in nuclear, particle, and solid-state physics. *Rev. Mod. Phys.* **87**, 137–163 (2015).

11. G. Baskaran, S. Mandal, R. Shankar, Exact results for spin dynamics and fractionalization in the Kitaev Model. *Phys. Rev. Lett.* **98**, 247201 (2007).

12. J. Knolle, D. L. Kovrizhin, J. T. Chalker, R. Moessner, Dynamics of a two-dimensional quantum spin liquid: Signatures of emergent Majorana fermions and fluxes. *Phys. Rev. Lett.* **112**, 207203 (2014).

13. J. Nasu, M. Udagawa, Y. Motome, Thermal fractionalization of quantum spins in a Kitaev model: Temperature-linear specific heat and coherent transport of Majorana fermions. *Phys. Rev. B*. **92**, 115122 (2015).

14. Y. Yamaji *et al.*, Clues and criteria for designing a Kitaev spin liquid revealed by thermal and spin excitations of the honeycomb iridate $Na_2IrO_3$. *Phys. Rev. B*. **93**, 174425 (2016).

15. J. Yoshitake, J. Nasu, Y. Motome, Fractional spin fluctuations as a precursor of quantum spin liquids: Majorana dynamical mean-field study for the Kitaev model. *Phys. Rev. Lett.* **117**, 157203 (2016).

16. B. J. Kim *et al.*, Novel $J_{eff}$=1/2 Mott state induced by relativistic spin-orbit coupling in $Sr_2IrO_4$. *Phys. Rev. Lett.* **101**, 076402 (2008).

17. S. H. Chun *et al.*, Direct evidence for dominant bond-directional interactions in a honeycomb lattice iridate $Na_2IrO_3$. *Nat. Phys.* **11**, 462–466 (2015).

18. S. K. Choi *et al.*, Spin waves and revised crystal structure of honeycomb iridate $Na_2IrO_3$. *Phys. Rev. Lett.* **108**, 127204 (2012).

19. F. Ye, S. Chi, H. Cao, B. C. Chakoumakos, Direct evidence of a zigzag spin-chain structure in the honeycomb lattice: A neutron and x-ray diffraction investigation of single-crystal $Na_2IrO_3$. *Phys. Rev. B*. **85**, 180403 (2012).

20. K. W. Plumb, J. P. Clancy, L. J. Sandilands, V. V. Shankar, α−$RuCl_3$: A spin-orbit assisted Mott insulator on a honeycomb lattice. *Phys. Rev. B*. **90**, 041112(R) (2014).

21. A. Koitzsch *et al.*, $J_{eff}$ description of the honeycomb Mott insulator α−$RuCl_3$. *Phys. Rev. Lett.* **117**, 126403 (2016).

22. L. J. Sandilands, Y. Tian, K. W. Plumb, Y.-J. Kim, K. S. Burch, Scattering continuum and possible fractionalized excitations in α-RuCl$_3$. *Phys. Rev. Lett.* **114**, 147201 (2015).

23. A. Banerjee *et al.*, Proximate Kitaev quantum spin liquid behaviour in a honeycomb magnet. *Nat. Mater.* **15**, 733–740 (2016).

24. R. D. Johnson *et al.*, Monoclinic crystal structure of α-RuCl$_3$ and the zigzag antiferromagnetic ground state. *Phys. Rev. B*. **92**, 235119 (2015).

25. H. B. Cao *et al.*, Low-temperature crystal and magnetic structure of α−RuCl$_3$. *Phys. Rev. B*. **93**, 134423 (2016).

26. S. M. Winter, Y. Li, H. O. Jeschke, R. Valenti, Challenges in design of Kitaev materials: Magnetic interactions from competing energy scales. *Phys. Rev. B*. **93**, 214431 (2016).

27. R. Yadav *et al.*, Kitaev exchange and field-induced quantum spin-liquid states in honeycomb α-RuCl$_3$. *Sci Rep*. **6**, 37925 (2016).

28. S. Y. Park *et al.*, http://arxiv.org/abs/1609.05690v1 (2016).

29. See supplementary materials for details.

30. A. Banerjee *et al.*, http://arxiv.org/abs/1609.00103v1 (2016).

31. A. Catuneanu, Y. Yamaji, G. Wachtel, H.-Y. Kee, Y. B. Kim, http://arxiv.org/abs/1701.07837v1 (2017).

32. M. Bouvier, P. Lethuillier, D. Schmitt, Specific heat in some gadolinium compounds. I. Experimental. *Phys. Rev. B*. **43**, 13137–13144 (1991).

33. G. Xu, Z. Xu, J. M. Tranquada, Absolute cross-section normalization of magnetic neutron scattering data. *Rev. Sci. Instrum.* **84**, 083906 (2013).

34. R. A. Ewings *et al.*, Horace: Software for the analysis of data from single crystal spectroscopy experiments at time-of-flight neutron instruments. *Nucl. Instrum. Methods Phys. Res. Sect. A*. **834**, 132–142 (2016).

35. J. C. Wasse, P. S. Salmon, Structure of molten ScCl$_3$ and ScI$_3$ studied by using neutron diffraction. *J. Phys.: Condens. Matter*. **11**, 2171–2177 (1999).

36. S. W. Lovesey, *Theory of Neutron Scattering from Condensed Matter, Volume 2* (Oxford University Press, 1986).

37. H. Ebert *et al.*, The Munich SPR-KKR package, version 6.3, (2012).

38. J. Perdew, K. Burke, M. Ernzerhof, Generalized gradient approximation made simple. *Phys. Rev. Lett.* **77**, 3865–3868 (1996).

39. S. Sinn *et al.*, Electronic structure of the Kitaev material α-RuCl$_3$ probed by photoemission and inverse photoemission spectroscopies. *Sci Rep*. **6**, 39544 (2016).

40. L. J. Sandilands, Y. Tian, A. A. Reijnders, H. S. Kim, Spin-orbit excitations and electronic structure of the putative Kitaev magnet α−RuCl$_3$. *Phys. Rev. B*. **93**, 075144 (2016).

41. X. Zhou *et al.*, Angle-resolved photoemission study of the Kitaev candidate α−RuCl$_3$. *Phys. Rev. B*. **94**, 161106 (2016).

42. P. J. Brown, "Magnetic form factors" in *International Tables for Crystallography, Volume C*, E. Prince, Ed. (Springer Science & Business Media, 2004), chap. 4.4.5.

43. N. G. Parkinson *et al.*, Study of the magnetic interactions in $Ba_2PrRu_{1-x}Cu_xO_6$ using neutron powder diffraction. *J. Mater. Chem.* **15**, 1375–1383 (2005).

44. A. Gukasov, M. Braden, R. J. Papoular, S. Nakatsuji, Y. Maeno, Anomalous spin-density distribution on oxygen and Ru in $Ca_{1.5}Sr_{0.5}RuO_4$: Polarized neutron diffraction study. *Phys. Rev. Lett.* **89**, 087202 (2002).

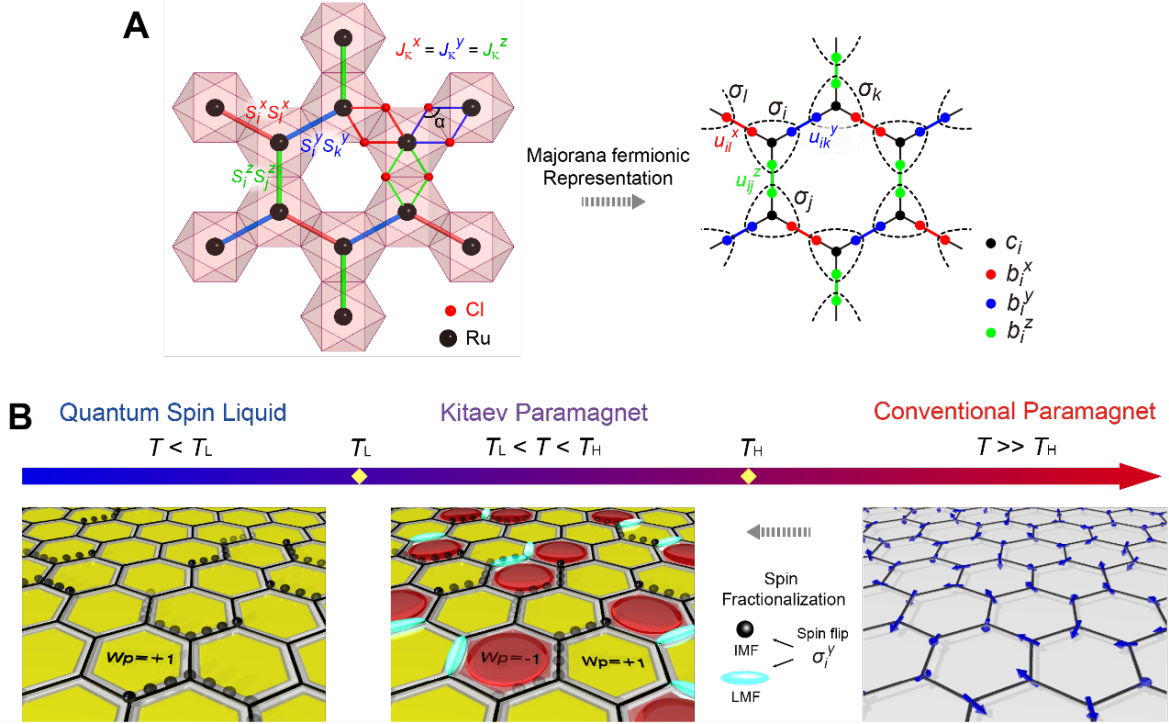

**Fig. 1**. **Kitaev bonding geometry and cartoon of emergent Majorana fermions.** (**A**) Local $Ru^{3+}$ ($J_{eff}$ = ½; $4d^5$) hexagon structure formed by the edge-shared $RuCl_6$ octahedra, in the layered honeycomb material α-$RuCl_3$. Two 94° Ru-Cl-Ru superexchange paths lead to the Kitaev interactions $J_K^\gamma S_i^\gamma S_j^\gamma$ between two magnetic spins on adjacent $i$ and $j$ sites, and the three different links denoted with $\gamma$ (= $x, y, z$) contribute isotropic $J_K$ in the rhombohedral crystal structure. The Paul spin operators can be represented by $\sigma_i^\gamma = ib_i^\gamma c_i$ in terms of fractionalized Majorana fermions ($c_i, b_i^x, b_i^y, b_i^z$) in an extended Hilbert space. $c_i$ denotes the itinerant Majorana fermion, and a product of the bond operators $u_{ij}^\gamma = ib_i^\gamma b_j^\gamma$ around the hexagon results in the $Z_2$ gauge flux $w_p = \pm 1$ in the Kitaev lattice. (**B**) (Left) In very low-temperature $T < T_L$, the $Z_2$-fluxes are almost frozen to the quantum spin liquid ground state with all $w_p = +1$ (yellow hexagon) and only low-energy itinerant Majorana fermions (black balls) are thermally activated. As temperature increases, the spin excitations are thermally fractionalized into itinerant and localized (cyan ovals) Majorana fermions. (Middle) In the intermediate temperature $T_L < T < T_H$, the fluxes are disordered to excite the localized Majorana fermions involving the flux-flips ($w_p = -1$, red hexagons) and the itinerant Majorana fermions on the vertices move in a coherent manner. As temperature crosses over $T_H$, the nearest-neighbor spin-spin correlation is diminished, and in high temperature $T \gg T_H$ (Right), the system becomes a conventional paramagnet.

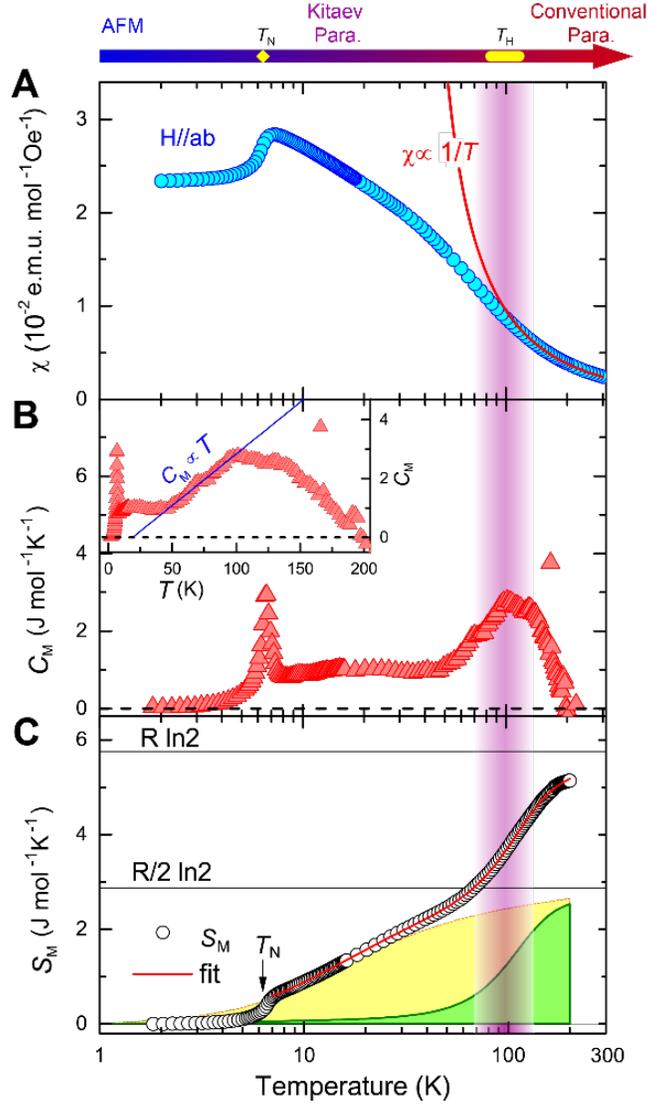

**Fig. 2. Thermodynamic signatures of spin fractionalization.** (**A**) Temperature dependent static magnetic susceptibility of α-RuCl$_3$ plotted in a semi-log scale for H//*ab*. The susceptibility deviates from the Curie-Weiss behavior (solid red line) below $T$ = 140 K. The low-temperature kink at $T_N$ = 6.5 K indicates a zigzag-type AFM order. (**B**) Magnetic specific heat $C_M$ obtained by subtracting the lattice contribution in a semi-log scale (*29*). Besides the AFM peak at $T_N$, the broad bumps in $T_N \lesssim T \lesssim 50$ K and around $T = T_H \simeq 100$ K (vertical bar) are associated with excitations of localized and itinerant Majorana fermions, respectively. $C_M$ exhibits a *T*-linear dependence in the intermediate temperature 50 K $\lesssim T \lesssim T_H$ as shown in the inset, reflecting metal-like density of states of the itinerant Majorana fermions. The spike at 165 K is due to a structural phase transition. (**C**) Magnetic entropy change, integrated $C_M$, in the temperature range 2 K < $T$ < 200 K. The horizontal solid lines represent the expected total entropy change $R\ln2$ and its half value $(R/2)\ln2$. The solid red line is a sum of two phenomenological function fits based on the theoretical simulation, indicating that the entropy release is decomposed into two fermionic components (yellow and green shadings) as described in Supplementary Materials.

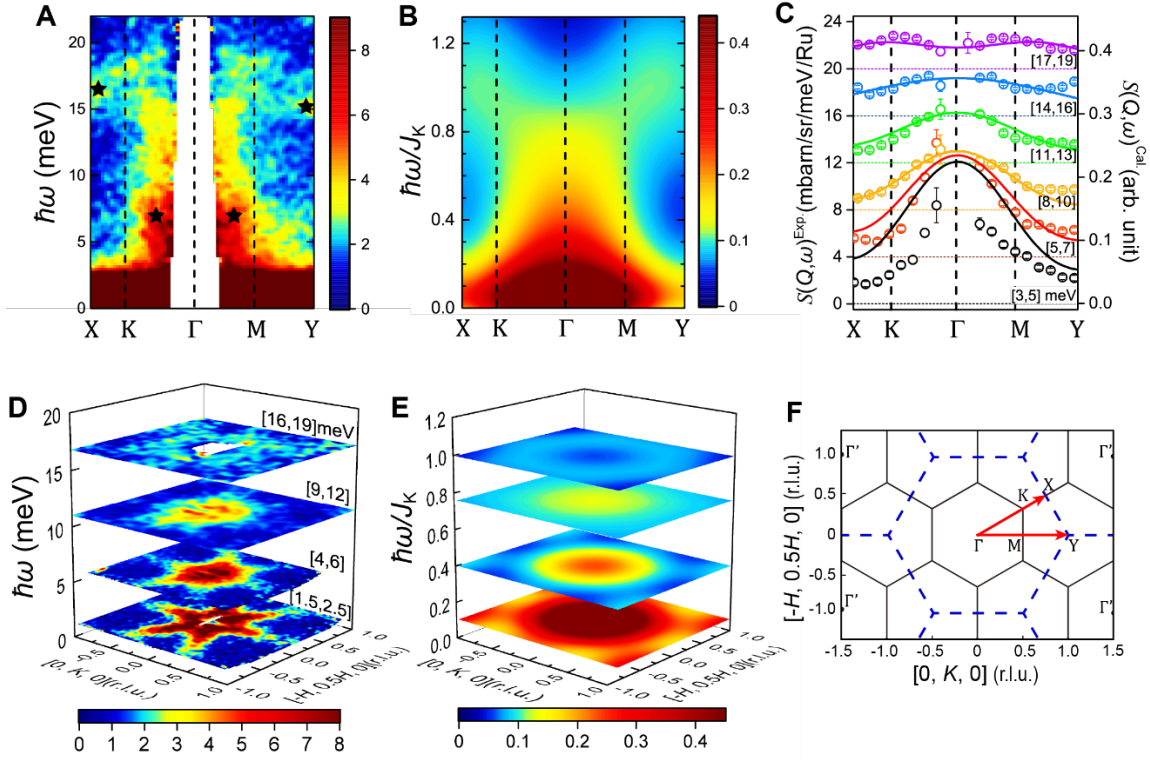

**Fig. 3. Magnetic excitation spectra of α-RuCl₃ compared with the theoretical calculations on a ferromagnetic Kitaev model.** (**A**) Neutron scattering function $S_{tot}(\mathbf{Q},\omega)$ at $T = 10$ K along the high symmetric line X-K-Γ-M-Y through Brillouin zone. The data were collected with an incoming neutron energy of $E_i = 31$ meV (MERLIN). The black stars mark phonons (*29*). (**B**) Calculated magnetic scattering function $S_{mag}(\mathbf{Q},\omega)$ for a ferromagnetic Kitaev model at $T = 0.06|J_K|$. (**C**) Constant-energy cuts integrated over the energy ranges [3,5], [5,7], [8,10], [11,13], [14,16], and [17,19] meV along the X-K-Γ and Γ-M-Y directions (left *y*-axis). The dashed lines guide vertical offsets. The solid lines present the theoretical calculations of the pure Kitaev model (right *y*-axis). (**D**) Constant-energy cuts in the (*hk*)-plane integrated over the energy ranges [1.5,2.5], [4,6] (LET, $E_i = 10$ meV), [9,12] (LET, $E_i = 22$ meV), and [16,19] meV (MERLIN, $E_i = 31$ meV). (**E**) Constant-energy cuts of the theoretical $S_{mag}(\mathbf{Q})$ in the Kitaev model for comparison. (**F**) The reciprocal honeycomb lattice in the $R\bar{3}$ space group. The X-K-Γ and Γ-M-Y directions are presented with the red arrows. The white regions in (**A**) and (**B**) mark the lack of detector coverage. The color bars in (**A**) and (**D**) are represented in the unit of mbarn·sr⁻¹·meV⁻¹ per Ru. The calculations presented in (**B**) and (**E**) are dimensionless, with the scale given by the color bar.

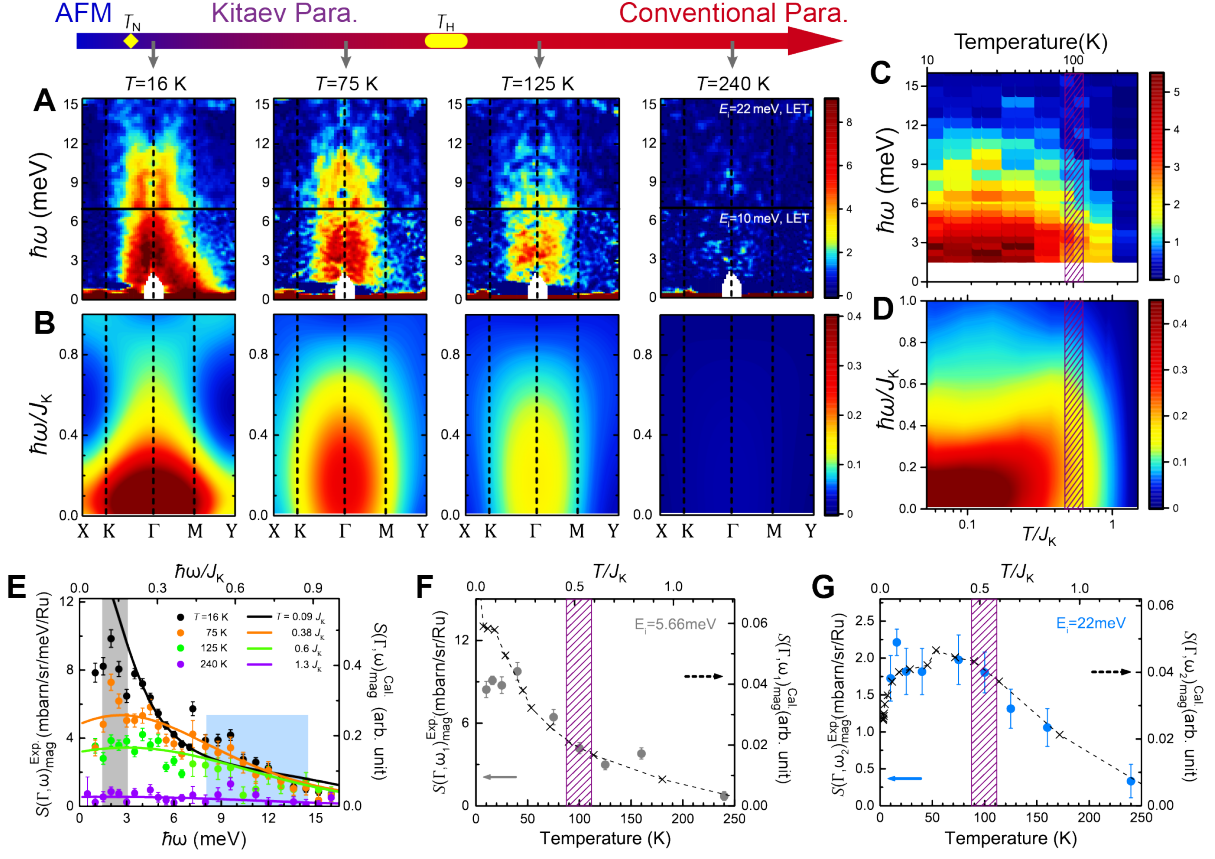

**Fig. 4. Evolution of the two Majorana fermion excitations.** **(A)** Magnetic scattering function $S_{mag}(Q,\omega)$ at $T$ = 16, 75, 125, and 240 K. The two data sets with an incoming neutron energy of $E_i$ = 22 meV (upper panel) and 10 meV (down panel) are combined together. The white regions mark the lack of detector coverage. **(B)** Calculated $S_{mag}(Q,\omega)$ at $T$ = 0.09, 0.375, 0.69, and 1.32$|J_K|$ with $J_K$ = -16.5 meV for comparison with the experimental data. **(C)-(D)** Comparison of contour plot of the experimental $S_{mag}(\omega)$ and the calculated $S_{mag}(\omega)$ at the $\Gamma$-point in the temperature-energy plane. **(E)** $S_{mag}(\Gamma,\omega)$ at $T$ = 16 K (black circles), 75 K (yellow circles), 125 K (green circles), and 240 K (blue circles) as a function of energy. The calculated $S_{mag}(\Gamma,\omega)$ (the solid lines) are presented together for comparison. **(F)-(G)** Temperature dependence of the integrated $S_{mag}(\Gamma,\omega)$ over the energy range $\omega_1$ = [1.5, 3] meV (grey circles) and $\omega_2$ = [8, 14.5] meV (blue circles). Both energy ranges are marked with grey and blue areas in e, respectively. The cross symbols represent the calculated results of the integrated $S_{mag}(\Gamma,\omega)$, and the dashed lines represent the linear interpolations. The areas of diagonal stripes in **(C)-(D)** and **(F)-(G)** indicate the high-$T$ crossover at $T_H$. The color bars in a and c are represented in the unit of mbarn·sr$^{-1}$·meV$^{-1}$ per Ru. The calculations presented in **(B)** and **(D)** are dimensionless, with the scale given by the color bar. In **(E)-(G)**, measured and calculated $S_{mag}$ refer to the left and right $y$-axes, respectively.

**Supplementary Materials:**

**Materials and Methods**

Crystal growth

High-quality single crystals of α-RuCl$_3$ and their isostructural counterpart ScCl$_3$ were grown by a vacuum sublimation method. A commercial RuCl$_3$ (ScCl$_3$) powder (Alfa-Aesar) was thoroughly ground, and dehydrated in a quartz ampoule for a day. The ampoule was sealed in vacuum and placed in a temperature gradient furnace. The temperature of the RuCl$_3$ (ScCl$_3$) powder is set at 1080 °C (900 °C). After dwelling for 5 hours the furnace is cooled to 650 °C (600 °C) at a rate of -2 °C per hour. We obtained α-RuCl$_3$ (ScCl$_3$) crystals black colored (transparent) with shinny surfaces. Electron dispersive X-ray measurements confirmed the stoichiometry of the Ru(Sc):Cl = 1:3 ratio for the crystals.

Magnetic susceptibility and specific heat measurement

Magnetic susceptibility measurements were performed using a commercial superconducting quantum interference device (SQUID) (Quantum Design, model: MPMS-5XL). A single domain crystal (3 x 3 x 1 mm$^3$, 20 mg) was chosen for the measurements under an external magnetic field parallel to the *ab*-plane. Specific heat $C_P$ was measured by using a conventional calorimeter of the Quantum Design Physical Property Measurement System (model: PPMS DynaCool) in a temperature range of $T$ = 1.8 - 300 K. The magnetic specific heat $C_M$ of α-RuCl$_3$ was determined by subtracting the lattice contribution, which is supposed to be equivalent to the specific heat of the iso-structural non-magnetic ScCl$_3$ with effective Debye temperature scaling (*32*).

Inelastic neutron scattering

Inelastic neutron scattering data were collected by using time-of-flight spectrometers MERLIN (high intensity) and LET (high resolution) at the ISIS spallation neutron source, the Rutherford Appleton Laboratory in the United Kingdom. Total 46 pieces (~1.35 g) of α-RuCl$_3$ single crystals for MERLIN, and 153 pieces (~ 5.1 g) for LET were prepared, and co-aligned with crystallographic *c*-axis surface normal on aluminum plates, resulting in a mosaic within 3° (Fig. S1). The samples were mounted in a liquid helium cryostat for temperature control ranging from 1.5 K to 270 K. Due to highly two-dimensional structure of α-RuCl$_3$, magnetic correlations between honeycomb layers are extremely weak and insensitive. Therefore, crystals are aligned with the c-axis parallel to the incident neutron beam, so that the area detector measures the energy spectrum over 2D *q*-space of *hk*-plane. To observe intensity at Γ-point (LET measurement), we rotated the crystal by 30 degrees to the incident beam direction, so that filled the blank region of beam mask.

Data were obtained with the incident neutron energy set to $E_i$ = 5.66, 10, 22 (LET), and 31 meV (MERLIN). With incoherent neutron scattering intensity measured from a vanadium standard sample, all data were normalized and converted to the value of neutron scattering function $S_{\text{tot}}(\mathbf{Q}, \omega)$ which is proportional to the differential neutron cross-section $\frac{d^2\sigma}{d\Omega dE}$ and the ratio of incident to scattered neutron wave-vector $k_i/k_f$ (*33*),

$$S_{\text{tot}}(\mathbf{Q}, \omega) \sim \frac{k_i}{k_f} \frac{d^2\sigma}{d\Omega dE}. \tag{S1}$$

Since $S_{\text{tot}}(\mathbf{Q}, \omega)$ contains both the nuclear and magnetic scattering contributions, the magnetic scattering function $S_{\text{mag}}(\mathbf{Q}, \omega)_T$ at temperature *T* in Fig. 4 is extracted from $S_{\text{tot}}(\mathbf{Q}, \omega)_T$ after subtraction of the scaled $S_{\text{tot}}(\mathbf{Q},\omega)_{T_0=290\text{ K}}$ with the Bose factor correction $n(T)/n(T_0) = (1-e^{-\hbar\omega/k_BT_0})/(1-e^{-\hbar\omega/k_BT})$, which represents the approximate phonon contribution in the experiment.

$$S_{\text{mag}}(\mathbf{Q},\omega)_T \approx S_{\text{tot}}(\mathbf{Q},\omega)_T - \frac{n(T)}{n(T_0)} S_{\text{tot}}(\mathbf{Q},\omega)_{T_0} \tag{S2}$$

All of data processes including Bose factor correction and projection of the scattering function along appropriate directions were performed by using the HORACE software, which is published by ISIS (*34*).

Calculation of the magnetic scattering function

The calculation of $S_{\text{mag}}(\mathbf{Q},\omega)_T$ is performed by using the CDMFT + continuous-time QMC method as described in Ref. 14. The Bose factor correction of the Eqn. (S2) is also applied to the simulation results for quantitative comparison with the experimental results in Fig. 4. All calculated results include the magnetic form factor of $Ru^{3+}$ ion, which is obtained by using the density functional theory method considering solid state effects in α-RuCl₃ as described in Supplementary Text.

**Supplementary Text**

Magnetic specific heat and entropy of α-RuCl₃

The specific heat $C_P$ of α-RuCl₃ consists of a lattice contribution $C_L$ and a magnetic contribution $C_M$. To estimate $C_L$, we measure the specific heat of iso-structural non-magnetic ScCl₃ (*35*), and then scale it by the renormalized Debye temperature ratio of $\Theta_D(\text{RuCl}_3)/\Theta_D(\text{ScCl}_3)$ (*32*). Figure S2 shows $C_P$ and the estimated $C_L$ of α-RuCl₃. The $C_M$, which is displayed in Fig. 2B of the main text, is obtained by subtracting $C_L$ from $C_P$.

The magnetic entropy $S_M(T) = \int C_M/T dT$ shows two-stage entropy release in addition to 19 % of the total entropy held by the antiferromagnetic order at $T_N = 6.5$ K (*28*) as shown in Fig. 2C of the main text. The low and high temperature stages correspond to thermal excitations of the localized and itinerant Majorana fermions (MFs), respectively. We decompose the stepwise $S_M(T)$ via fitting with the sum of two phenomenological Schottky-like functions introduced by Yamaji *et al.* (*14*),

$$S_M = \sum_{a=L,H} \frac{\rho_a/2}{1+\exp\left[\left(\frac{\beta_a+\gamma_a T_a/T}{1+T_a/T}\right)\ln\left(\frac{T_a}{T}\right)\right]}, \tag{S3}$$

which well reproduces the simulated entropy for the Kitaev-Heisenberg model. Here L and H are indexes for the respective localized and itinerant MFs, and $\rho_a$ is the weight of each entropy under a constraint of $\rho_L + \rho_H = 2$. $\beta_a$ and $\gamma_a$ correspond to exponents in the high- and low-temperature power behaviors with the crossover temperature $T_a$, respectively. Fitting was carried out in the temperature range, $T_N < T < 200$ K, and the fitting parameters are listed in the Table S1. The fitting result indicate that the entropy weight corresponding to each MF is almost equal as predicted in the Kitaev honeycomb model (*13, 14*). The $T_L$ value in the fit is rather large, likely due to the additional perturbing magnetic interactions (*14, 31*), which are expected to make considerable influence on the low energy MF excitation behaviors.

Neutron scattering function and phonon modes at $T = 10$ K

The neutron scattering function at temperature $T$ consists of both the nuclear and magnetic scattering contributions, which is expressed as functions of the momentum transfer $\mathbf{Q}$, energy transfer $\omega$,

$$S_{\text{tot}}(\mathbf{Q},\omega)_T = S_{\text{mag}}(\mathbf{Q},\omega)_T + S_{\text{ph}}(\mathbf{Q},\omega)_T. \tag{S4}$$

$S_{\text{mag}}(\mathbf{Q},\omega)_T$ representing the magnetic scattering from the sample is described as

$$S_{\text{mag}}(\mathbf{Q},\omega)_T \sim g^2 |f(Q)|^2 \int_{-\infty}^{\infty} dt\, e^{-i\omega t} \sum_{i,j} e^{i\mathbf{Q}\cdot\mathbf{r}_{ij}} \langle S_i^{\gamma}(t) S_j^{\gamma}(0)\rangle_T, \tag{S3}$$

with the g-factor $g$ and the magnetic form factor $f(Q)$ of $Ru^{3+}$ ion. $\langle S_i^{\gamma}(t) S_j^{\gamma}(0)\rangle_T$ ($\gamma = x, y, z$ in the Kitaev honeycomb model) presents the dynamical correlation function between the nearest neighbor spins at $i$- and $j$-th sites, which is involved in a spin-flip process by either thermal activation or neutron scattering at temperature $T$ (12, 15). $S_{\text{ph}}(\mathbf{Q},\omega)_T$ representing the nuclear (phonon) scattering from the sample and aluminium holder has a function form of

$$S_{\text{ph}}(\mathbf{Q},\omega)_T \sim \frac{n(T)}{\omega(\mathbf{q})} |F_N(\mathbf{G})|^2 \frac{|\mathbf{Q}\cdot\boldsymbol{\xi}|^2}{2M} \delta(\omega - \omega(\mathbf{q})), \tag{S5}$$

where the phonon wave-vector $\mathbf{q}$ and Bragg wave-vector $\mathbf{G}$ are related by $\mathbf{Q} = \mathbf{q} + \mathbf{G}$. $\boldsymbol{\xi}$ denotes the unit vector along the phonon polarization direction and $\omega(q)$ is the phonon frequency. $|F_N(\mathbf{G})|$ is the nuclear structure factor and $n(T) = 1/(1 - e^{-\hbar\omega/k_B T})$ is the Bose factor. Figure S3 displays $S_{\text{tot}}(\mathbf{Q},\omega)_{T=10K}$ (Fig. S3A) and $S_{\text{tot}}(\mathbf{Q},\omega)_{T=200K}$ (Fig. S3B) measured with the MERLIN spectrometer, which are dominated by the magnetic and phonon scattering, respectively. Both spectra follow the general features of magnetic and phonon scattering with $\mathbf{Q}$ and $T$ variations; $S_{\text{mag}}(\mathbf{Q},\omega)$ dominates at low $\mathbf{Q}$ and $T$ while $S_{\text{ph}}(\mathbf{Q},\omega)$ dominates at high $\mathbf{Q}$ and $T$ as confirmed in $S_{\text{tot}}(\mathbf{Q},\omega)_{T=200K} - S_{\text{tot}}(\mathbf{Q},\omega)_{T=10K}$ (Fig. S3C). While $S_{\text{tot}}(\mathbf{Q},\omega)_{T=10K}$ is dominated by magnetic scattering, it contains weak phonon modes deduced from Fig. S3C. Intensity spots around at X- and Y-point near 15 meV and along K-Γ-M direction near 7 meV becomes strong at $T$ = 200 K, which is the typical temperature-dependent phonon behavior due to the Bose factor. Those modes are marked with star symbols in Fig. 3A of the main text.

Temperature dependence of anisotropic low-energy excitations

Figure S4A shows the $Q$-distributions of low-energy excitations obtained from experimental $S_{\text{mag}}(\mathbf{Q},\omega)$ by integration in the energy range [1.5, 3] meV at several temperatures. The hexagram-type anisotropy appearing up to 40 K due to perturbing magnetic interactions (30, 31) becomes isotropic above ~ 50 K as expected in the Kitaev model. This behavior is supported from over-plots of line-cuts $S_{\text{mag}}(Q)$ along [0, 1, 0] and [1, 1, 0] directions (Fig. S4B), which shows that the discrepancy between two line-cuts at $T$ = 16 K reduces upon warming and disappears above ~ 50 K.

Magnetic form factor of $Ru^{3+}$

In the magnetic scattering, the magnetic form factor depends on the true magnetic spin density of the magnetic ion, for which the solid state effects are taken into account. In α-RuCl$_3$, the spin density of $Ru^{3+}$ 4$d$ electrons is naturally influenced by the interactions with the surrounding ligand ions such as the Ru 4$d$-Cl 2$p$ hybridization and the crystal field splitting. In order to obtain the proper magnetic form factor, we performed the first principles density functional theory (DFT) calculations. The magnetic form factor $f(Q)$ is calculated within dipole approximation which is valid in the limit of small wave vector transfer $Q$ (36).

$$f(Q) = \int \rho_{\text{sp}}(r) j_0(Qr) r^2 dr + \int \rho_{\text{orb}}(r)\{j_0(Qr) + j_2(Qr)\} r^2 dr, \tag{S6}$$

where $j_n(Qr)$'s are Bessel functions and $\rho_{\text{sp/orb}}(r)$ is spin/orbital density. The integration is performed inside the muffin-tin.

To get the ground state, the self-consistent band structure calculations were carried out, by using the spin polarized, full relativistic Korringa-Kohn-Rostocker Green function method, implemented in SPR-KKR package (*37*). For the exchange-correlation potential, generalize gradient approximation in the Perdew-Burke-Ernzerhof parametrization scheme is used (*38*). For the Brillouin zone integration, $23 \times 13 \times 8$ *k*-point mesh is used. To take account of the correlation effects of Ru 4*d* orbitals, we adopted DFT+U method in its atomic limit form, using the on-site Coulomb U = 3.0 eV and exchange J = 0.4 eV. The band gap of about 1.5 eV is obtained for the converged ground state, which describes the observed gap reasonably well, although experimental gap data are rather widely spread ranging from 1 eV to 1.9 eV (*39-41*). The magnetic moment is estimated to be ~ 0.67 $\mu_B$ per Ru, in a good agreement with the observed value 0.73 $\mu_B$ (*28*). Note that the Ru form factor is available only for $Ru^{1+}$ and $Ru^{5+}$ in literature (*42-44*). In Fig. S5, the magnetic form factor of $Ru^{3+}$ in $\alpha$-$RuCl_3$ calculated by DFT (blue line) is compared with existing magnetic form factors with the different valence state.

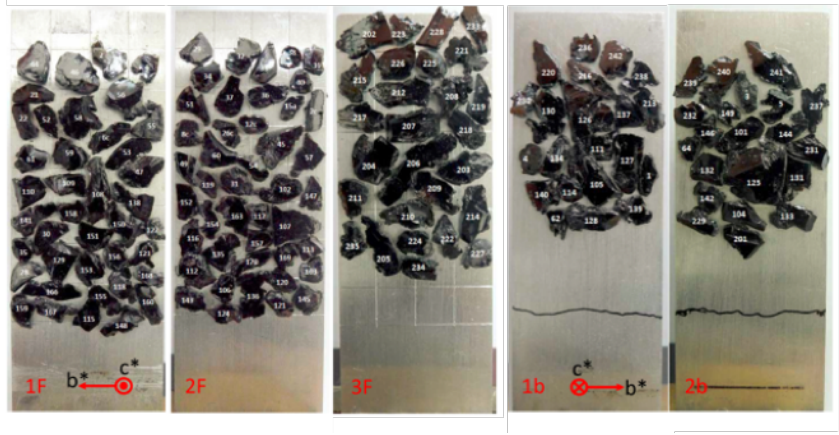

**Fig. S1.**

Photograph of co-aligned samples for the inelastic neutron scattering measurement. Total 153 pieces (total mass is 5.1 g) of α-RuCl$_3$ were attached on three aluminium plates by using a cytop glue.

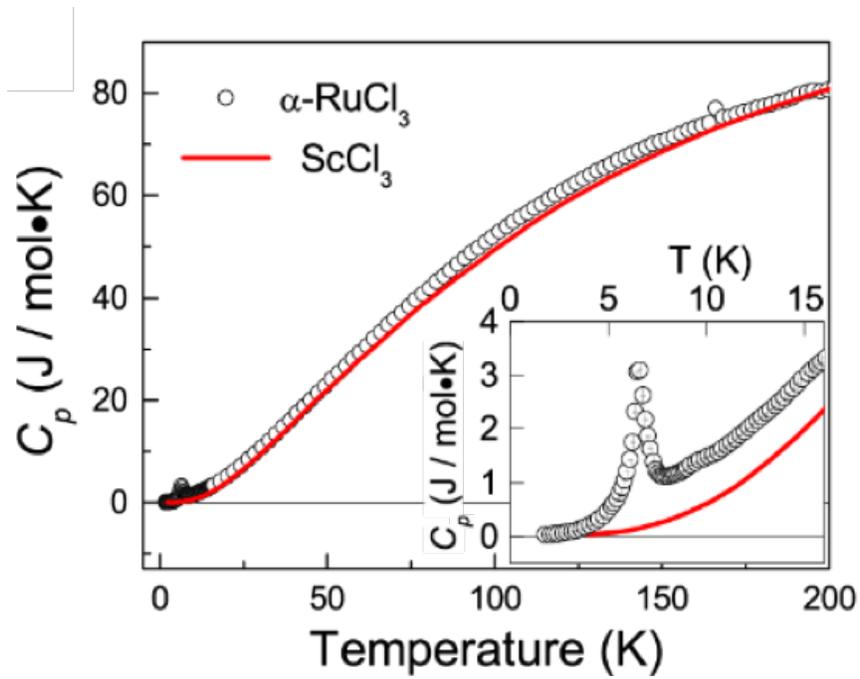

**Fig. S2**
Specific heats $C_P$ (circle) of α-RuCl$_3$ and its lattice contribution $C_L$ (red line) estimated from ScCl$_3$. The zero baseline is marked with a black dashed line. Inset shows the magnified specific heats below 16 K. The $C_P$ of α-RuCl$_3$ exhibits the antiferromagnetic transition peak at $T_N$ = 6.5 K.

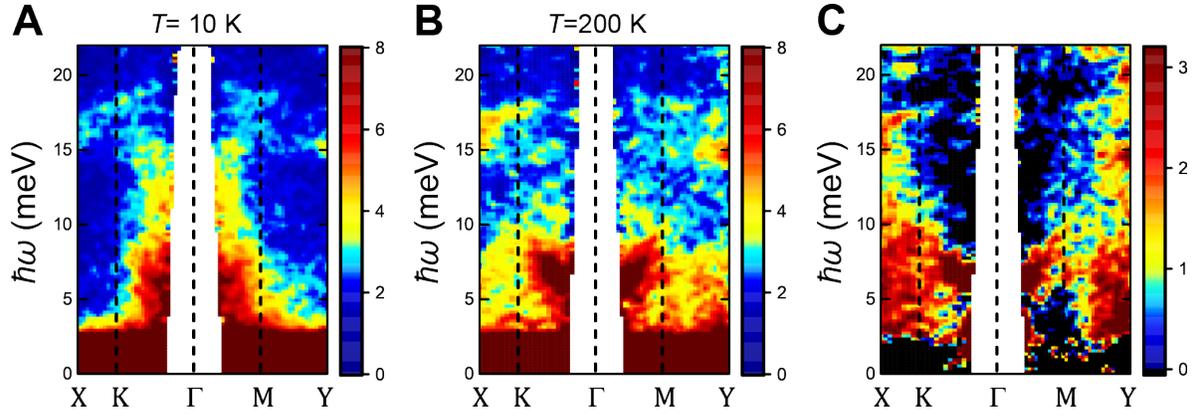

**Fig. S3**

Neutron scattering function $\mathcal{S}_{\text{tot}}(\mathbf{Q}, \omega, T)$ measured at MERLIN spectrometer with an incident neutron energy $E_i = 31$ meV. The direct beam mask is shaded by white rectangular area around at Γ-point. **(A)** $\mathcal{S}_{\text{tot}}(\mathbf{Q}, \omega, T)$ at $T = 10$ K. **(B)** $\mathcal{S}_{\text{tot}}(\mathbf{Q}, \omega, T)$ at $T = 200$ K. **(C)** $\mathcal{S}_{\text{tot}}(\mathbf{Q}, \omega, T = 200 \text{ K}) - \mathcal{S}_{\text{tot}}(\mathbf{Q}, \omega, T = 10 \text{ K})$.

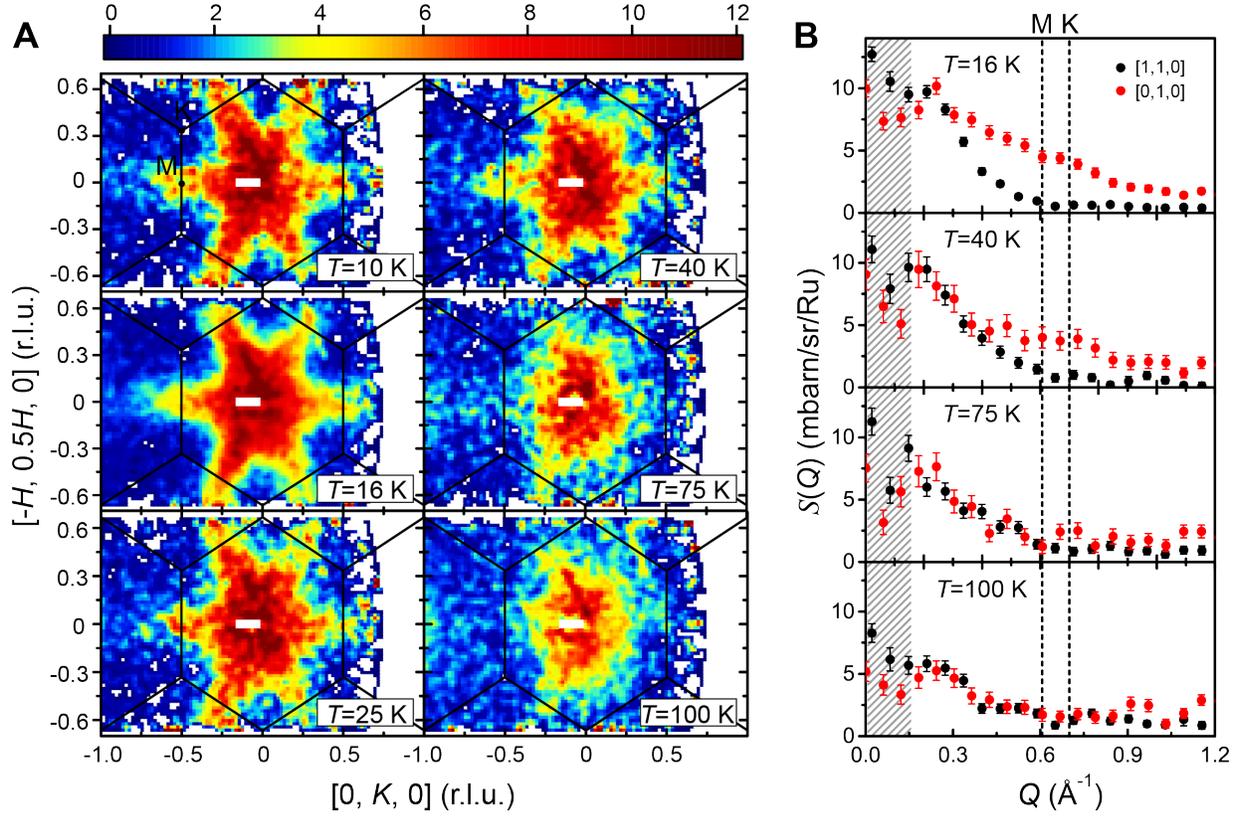

**Fig. S4**

(A) Contour plots of $S_{mag}(\mathbf{Q})$ integrated over the energy range [1.5, 3] meV in the *hk*-plane at $T$=10, 16, 25, 40, 75, and 100 K. The data were collected using the LET time-of-flight spectrometer with an incoming neutron energy of $E_i$ = 5.66 meV. The white region near Brillouin Zone center marks the lack of detector coverage and solid lines indicate Brillouin zone boundaries. (B) Line-cuts $S_{mag}(Q)$ along [0, 1, 0] (red circles) and [1, 1, 0] (black circles) direction at $T$ = 16, 40, 75, and 100 K. The shaded region is affected by direct beam mask.

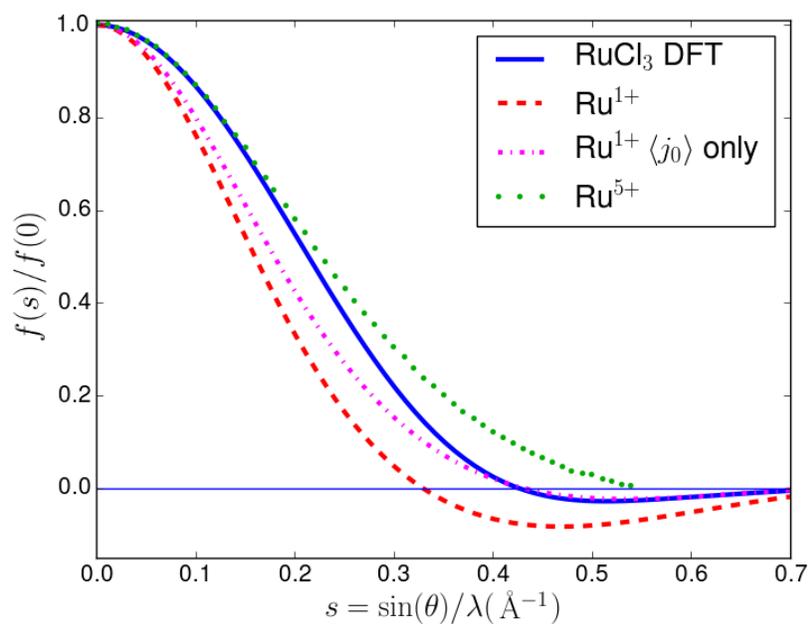

**Fig. S5**

The magnetic form factor of $Ru^{3+}$ in α-$RuCl_3$ obtained from the DFT calculations (blue line). Magnetic form factors of Ru ions with different valences and configurations are drawn together for comparison: $Ru^{1+}$ ion (red dashed line) and its spin only contribution (dash-dot line), and $Ru^{5+}$ (green dotted line).

**Table S1.**

Fitting parameters obtained from the Eqn. (S1).

| parameter | Value | δ |
|---|---|---|
| $\rho_H$ | 0.92 | 0.07 |
| $\rho_L$ | 1.08 | 0.07 |
| $\beta_H$ | 6.04 | 0.23 |
| $\beta_L$ | 0.72 | 0.05 |
| $\gamma_H$ | 1.11 | 0.76 |
| $\gamma_L$ | 1.55 | 0.17 |
| $T_H$ | 101.28 | 3.05 |
| $T_L$ | 22.22 | 0.54 |